\documentstyle[12pt]{article} 
\newcommand{\be}{\begin{equation}}
\newcommand{\nn}{\nonumber}
\newcommand{\bea}{\begin{eqnarray}}
\newcommand{\eea}{\end{eqnarray}}
\newcommand{\ba}{\begin{array}}
\newcommand{\ea}{\end{array}}
\newcommand{\ee}{\end{equation}}

\expandafter\ifx\csname mathbbm\endcsname\relax

\else

\fi \textheight 22cm \textwidth 15cm \topmargin 1mm \oddsidemargin
5mm \evensidemargin 5mm

\newcommand{\beas}{\begin{eqnarray*}}
\newcommand{\eeas}{\end{eqnarray*}}
\newcommand{\bes}{\begin{equation*}}
\newcommand{\ees}{\end{equation*}}

\newcommand{\lf}{\left}
\newcommand{\ri}{\right}
\newcommand{\f}{\frac}

\def\i2           {\mbox{$\frac{i}{2}$}}

\def\al           {\alpha}

\def\bet           {\beta}

\def\lat            {\tilde{\la}}
\def\io            {{\rm Im}\, \Omega}
\def\ro            {{\rm Re}\, \Omega}

\def\del           {\delta}

\def\ep           {\epsilon}

\def\et           {\eta}

\def\ga           {\gamma}

\def\la           {\lambda}

\def\om           {\omega}

\def\ph           {\phi}

\def\si           {\sigma}
\def\Si           {\Sigma}

\def\th{\theta}

\def\eh  {{\hat e}}
\def\we {{\wedge}}

\def\eh    {{\hat{e}}} 
\begin{document}

\begin{titlepage}
\vspace*{20mm}
\begin{center}
{\LARGE \bf{{New Compactifications of \\
Eleven Dimensional Supergravity}}}\\ 

\vspace*{1cm}
\vspace*{20mm} \vspace*{1mm} {Ali Imaanpur}

\let\thefootnote\relax\footnotetext{Email: aimaanpu@theory.ipm.ac.ir}

\vspace*{1cm}
  
{\it Department of Physics, School of Sciences\\ 
Tarbiat Modares University, P.O.Box 14155-4838, Tehran, Iran}

\vspace*{1mm}

\vspace*{1cm}

\end{center}

\begin{abstract}

Using canonical forms on $S^7$, viewed as an $SU(2)$ bundle over $S^4$, we introduce consistent 
ans\"atze for the 4-form field strength of eleven-dimensional supergravity and rederive the known 
squashed, stretched, and the Englert solutions. Further, by rewriting the metric of $S^7$ as a $U(1)$ bundle over 
${\bf CP}^3$, we present yet more general ans\"atze. As a result, we find a new compactifying solution of the type 
$AdS_5\times {\bf CP}^3$, where ${\bf CP}^3$ is stretched along its $S^2$ fiber. We also 
find a new solution of $AdS_2\times H^2\times S^7$ type in Euclidean space. 

\end{abstract}

\end{titlepage}

\section{Introduction}
Eleven-dimensional supergravity solutions have been extensively studied in 1980's. Among these, the 
Freund-Rubin solution \cite{RUB} was the simplest one as it included a 4-form field strength with 
components only along the $AdS$ direction. Then, attentions were turned to possible solutions 
with nonvanishing components along the compact directions. Englert was the first to construct such a 
solution; $AdS_4\times S^7$ with the round metric on $S^7$ \cite{ENG}. Later, the so-called squashed 
solutions with non-standard Einstein metric on $S^7$ were found \cite{AWA, DUF}. Here, $S^7$ is considered as 
an $SU(2)$ bundle over $S^4$ and squashing corresponds to rescaling the metric along the fiber. For a 
specific value of the squashing parameter the metric turns out to be Einstein. 

In constructing the Englert type solutions Killing spinors play a significant role. 
Killing spinors are also required for having supersymmetric solutions \cite{BAI, ROM, SOR, SOR2, ENGTO}. 
Alternatively, on compact manifolds with a bundle structure on a K\"ahler base, one can use the holomorphic 
top form and the K\"ahler form to write consistent ans\"atze for the 4-form field strength \cite{WAR}. 
Algebraic approaches have also been used to study the supergravity solutions \cite{VAS}. 

In the present work, however, instead of looking for Killing spinors we directly use canonical forms on $S^7$ to 
write a consistent ansatz for the 4-form field strength. In particular, this allows us to rederive the squashed,  stretched, and the Englert solutions in a unified scheme. There are some independent earlier works which also use 
canonical geometric methods \cite{GAU1, CAS}.  In Sec. 2, we consider $S^7$ as an $S^3$ bundle 
over $S^4$, and identify a natural basis of such forms in terms of the volume forms of the fiber and the base.   
We will see that a linear combination of these forms provides a suitable ansatz for the Maxwell equation, so that  
the field equations reduce to algebraic equations for the parameters of the ansatz. In Sec. 3, we rewrite the 
squashed metric of $S^7$ as a $U(1)$ bundle over ${\bf CP}^3$, where it appears as an $S^2$ bundle 
over $S^4$ with rescaled fibers. Moreover, in this form, we can introduce a different rescaling parameter for 
the $U(1)$ fibers. This enables us to provide more general ans\"atze. In Sec. 4, we consider a direct product of 
a 5 and 6-dimensional spaces and find a new compactifying solution of $AdS_5\times {\bf CP}^3$, in which 
${\bf CP}^3$ is stretched along its $S^2$ fiber. In Sec. 5, we discuss solutions in which the eleven dimensional 
space has a Euclidean signature and is a direct product of two 2-dimensional spaces and $S^7$. We find a solution of 
$AdS_2\times H^2\times S^7$ type, in which $H^2$ is a hyperbolic surface, and $S^7$ is stretched along its $U(1)$ 
fiber by a factor of $2$.

\section{Squashed solution revisited}
Let us start our discussion with the Freund-Rubin solution, for which the 4-form field strength has components 
only along the four dimensions
\be
F_4=\f{3}{8}R^3 \ep_4\, ,
\ee  
and the metric reads 
\be
ds^2= {R^2}(\f{1}{4}ds^2_{AdS_4} + ds^2_{S^7})\, .
\ee
The round metric on $S^7$ can be written as an $SU(2)$ bundle over $S^4$ \cite{AWA, DUF2}
\be
ds^2_{S^7}= \f{1}{4}(d\mu^2 +\f{1}{4} \sin^2 \mu\, \Sigma_i^2 + (\sigma_i -\cos^2{\mu}/{2}\ \Sigma_i)^2)\, ,\label{RO}
\ee
with $0\leq \mu \leq \pi$, and $\Sigma_i$'s and $\sigma_i$'s are two sets of left-invariant one-forms 
\bea
&&\Si_1=\cos \ga\, d\al +\sin \ga \sin \al\,  d\bet\, , \nn \\
&&\Si_2=-\sin \ga\, d\al +\cos \ga \sin \al\,  d\bet\, , \  \nn \\
&& \Si_3= d\ga +\cos \al\,  d\bet \, , \nn
\eea
where $0\leq \ga \leq 4\pi ,\, 0\leq \al \leq \pi ,\, 0\leq \bet \leq 2\pi$, and 
with a similar expression for $\si_i$'s. They satisfy the $SU(2)$ algebra
\be
d\Sigma_i=-\f{1}{2}\, \ep_{ijk}\, \Sigma_j \we \Sigma_k\, , \ \ \ \  d\sigma_i=-\f{1}{2}\, \ep_{ijk}\, \sigma_j \we \sigma_k\, ,
\label{SIG}
\ee
with $i,j,k,\ldots =1,2,3$. 

Squashing corresponds to modifying the round metric on $S^7$ as follows
\be
ds^2_{S^7}= \f{1}{4}(d\mu^2 +\f{1}{4} \sin^2 \mu\, \Sigma_i^2 + \la^2 (\sigma_i -\cos^2 {\mu}/{2}\, \Sigma_i)^2)\, ,
\ee
with $\la$ the squashing parameter. So, let us take the following ansatz for the 11d metric:
\be
ds^2= \f{R^2}{4}\lf(ds^2_4 \, +\, d\mu^2 +\f{1}{4} \sin^2 \mu\, \Sigma_i^2 + \la^2 
(\sigma_i -\cos^2 {\mu}/{2}\, \Sigma_i)^2\ri)\, ,\label{11MET}
\ee
and choose the orthonormal basis of vielbeins as
\be
e^0=\, d\mu\, , \ \ e^i=\f{1}{2} \sin \mu\, \Sigma_i\, ,\ \ \ \ \eh^i=\la (\sigma_i -\cos^2 {\mu}/{2}\, \Sigma_i)\, .
\ee
Further, in order to construct our ansatz in the next section we need to evaluate the exterior derivatives of the vielbeins 
\bea
de^i &=&\cot \mu\, e^0\we e^i -\f{1}{\sin \mu}\ep_{ijk} e^j\we e^k  \label{DIF1} \\
d\eh^i &=& \la\, e^0\we e^i +\f{1}{2}\ep_{ijk}\lf(\la\, e^j\we e^k-\f{1}{\la} \eh^j\we \eh^k -2\lf(\f{1+\cos \mu}{\sin \mu}\ri) 
e^j\we\eh^k \ri)\, , \label{DIF2}
\eea
where use has been made of (\ref{SIG}).

\subsection{The ansatz}
Let us now introduce $\om_3$, the volume element of the fiber $S^3$:
\be
\om_3 = \eh^1\we \eh^2 \we \eh^3\, ,
\ee
taking the derivative along with using (\ref{DIF2}), we obtain
\be
d\om_3 = \f{\la}{2}\lf( \ep_{ijk}\, e^0\we e^i \we \eh^j \we \eh^k +\, e^i\we e^j\we \eh^i\we \eh ^j\ri)\, .\label{OM3}
\ee
The Hodge dual reads
\be
*d\om_3={\la}\, \eh^i\we\, (e^0\we e^i +\f{1}{2}\ep_{ijk}\, e^j\we e^k )\, ,
\ee
so that using (\ref{DIF1}) and (\ref{DIF2}), we derive
\be
d*d\om_3 ={6\la^2}\, \om_4 - \f{1}{\la}\, d\om_3 \, ,\label{STA}
\ee
where
\be
\om_4= e^0\we e^1\we e^2\we e^3 \, ,
\ee
is the volume element of the base. Note that $\om_4$ is closed; $d\om_4 =0$. Further, since 
$d*\om_4=d\om_3$, for a linear combination of these two forms we have
\be
d*(\al\, \om_4 +\bet\, d\om_3)= {6\la^2 \bet}\, \om_4 +(\al-{\bet}/{\la})\, d\om_3 
\ee
namely, the subspace with a basis of $\om_4$ and $d\om_3$ is closed under $d*$ operation. This 
is exactly what we need to construct a consistent ansatz for the 4-form field strength. 

The above analysis shows that we can take the following ansatz:
\be
F_4= N\ep_4 +\al\, \om_4 +\bet\, d\om_3\, ,\label{ANS}
\ee
with $N$, $\al$, and $\bet$ constant parameters to be determined by field equations, 
also note that $dF_4 =0$.
Substituting this into the field equation\footnote{The star in this equation is the 
eleven-dimensional Hodge dual operation. In the rest of equations, it indicates 
the seven dimensional Hodge dual operation.} 
\be
d*_{11}F_4=-\f{1}{2}\, F_4\we\, F_4 \, ,\label{MAX}
\ee
we get
\be
\f{R^3}{8}d\lf( N\, \om_3\we \om_4 + \al \ep_4\we \om_3 +\bet \ep_4\we *d\om_3\ri) =-N \ep_4 \we\, (\al \om_4 +\bet d\om_3)\, ,
\ee
therefore, using (\ref{STA}), we must have
\be
{6\la^2}\, \bet =-\f{8N}{R^3}\, \al\, ,\ \ \ \ \ \ \al -\f{\bet}{\la }=-\f{8N}{R^3}\, \bet\, .
\label{N1}
\ee
A nontrivial solution exists if 
\be
\la \lf(\f{8N}{R^3}\ri)^2-\f{8N}{R^3}-6\la^3=0\, .\label{N2}
\ee
We will return to this equation after discussing the 
Einstein equations.

Now let us turn to the Einstein equations:
\be
R_{MN}=\f{1}{12}\, F_{MPQR}F_N^{\, \ PQR} -\f{1}{3\cdot 48}\, g_{MN}\ F_{PQRS}F^{PQRS}\, ,
\ee
where $M,N,P,\ldots =0,1,\ldots, 10$. With ansatz (\ref{ANS}), we can calculate the right hand side of the 
above equations: 
\bea
R_{\mu\nu}\! &=&\!\!\lf(\f{4}{R^2}\ri)^4 \lf(-\f{3!}{12}\, N^2-\f{4!}{3\cdot 48}(-N^2 +\al^2 +6\la^2\bet^2)\ri)
g_{\mu\nu}\label{E1}\, , \\
R_{\al\bet}\! &=&\!\!\lf(\f{4}{R^2}\ri)^4 \lf(\f{3!}{12}(\al^2+3\la^2\bet^2)-\f{4!}{3\cdot 48}
(-N^2 +\al^2 +6\la^2\bet^2)\ri) \del_{\al\bet} \, ,\label{E2} \\
R_{\hat{\al}\hat{\bet}}\!&=&\!\! \lf(\f{4}{R^2}\ri)^4 \lf(\f{3!}{12}(4\, \la^2\bet^2)-\f{4!}{3\cdot 48}
(-N^2 +\al^2 +6\la^2\bet^2)\ri) \del_{\hat{\al}\hat{\bet}}\, ,  \label{E3} 
\eea
with $\mu,\nu =0,\ldots,3,\, \al ,\bet =4,\ldots 7,$ and $\hat{\al}, \hat{\bet}=8,9,10$. 
Notice that different terms in our ansatz (\ref{ANS}) do not contract into each other. 
For the left hand side, on the other hand, the Ricci tensor of metric (\ref{11MET}) becomes
\bea
R_{\al\bet}&=&\lf(\f{4}{R^2}\ri)\lf(\f{3(2-\la^2)}{2}\ri)\, \del_{\al\bet}\, , \nn \\
R_{\hat{\al}\hat{\bet}}&=& \lf(\f{4}{R^2}\ri)\lf(\f{1+2\la^4}{2\la^2}\ri)\, \del_{\hat{\al}\hat{\bet}}\, ,
\eea
these are to be substituted on the left hand side of (\ref{E2}) and  (\ref{E3}). 

We can now solve (\ref{N1}) and (\ref{N2}) for $\bet$ and $N$, and then plug it into (\ref{E2}) and  (\ref{E3}). 
The two resulting equations can be solved  for $\la$ and $\al$. We get two types of solutions. Those with no internal flux: 
\be
\al =\bet=0\, , 
\ee
together with $\la^2=1$, which is the round sphere. Or, we can have $\la^2=1/5$, which corresponds to the squashed sphere solution. 
We also get solutions with fluxes
\be
\al^2=9/5\, ,\ \ \ \bet^2=9\, ,\ \ \ \ \la^2=1/5\, .\nn
\ee
For $\la={1}/{\sqrt{5}}$, $\al=-{3}/{\sqrt{5}} ,\ \bet=3$, and $N={3R^3}/(4\sqrt{5})$ we have a non-zero 
4-form field strength along $S^7$, and it represents the squashed $S^7$ with Einstein metric 
\be
R_{\al\bet}=\lf(\f{4}{R^2}\ri)\f{27}{10}\, \del_{\al\bet}\, , \ \ \ \ 
R_{\hat{\al}\hat{\bet}}=\lf(\f{4}{R^2}\ri)\f{27}{10}\, \del_{\hat{\al}\hat{\bet}}\, .
\ee
The above solution, the so-called squashed solution with torsion,  was obtained in 1980's using the covariantly 
constant spinors of the squashed sphere without torsion \cite{BAI, ENGTO}.
We can also take $\la={-1}/{\sqrt{5}}$, $\al=-{3}/{\sqrt{5}} ,\ \bet=-3$, and $N={-3R^3}/(4\sqrt{5})$  instead, this is the skew-whiffed squashed 
solution.  

\section{${\bf CP}^3$ as an $S^2$ bundle over $S^4$}
In the previous section the metric of $S^7$ was written as an $S^3$ bundle over $S^4$. It is also possible to write the 
metric as a $U(1)$ bundle over ${\bf CP}^3$. On the other hand, it is observed that ${\bf CP}^3$ itself can be written as an $S^2$ 
bundle over $S^4$. In this form one can construct a family of  homogeneous metrics by rescaling the fibers. 
In fact, we can see that the metric (\ref{RO}) can be rewritten as a $U(1)$ bundle over such a deformed 
${\bf CP}^3$ \cite{HOPF, FONT}.  
First note that\footnote{For the sake of clarity, we will set $R^2=4$ from now on.} 
\bea
ds^2_{S^7}&=& d\mu^2 +\f{1}{4} \sin^2 \mu\, \Sigma_i^2 +\la^2(\sigma_i -\cos^2{\mu}/{2}\ \Sigma_i)^2\nn \\
&=& d\mu^2 +\f{1}{4} \sin^2 \mu\, \Sigma_i^2 +\la^2(d\tau-A)^2+\la^2 (d\th -\sin \ph A_1+\cos \ph A_2)^2 \nn \\
&+&\la^2 \sin^2 \th\, (d\ph -\cot \th(\cos \ph A_1 +\sin \ph A_2)+A_3)^2\, , \label{NEE}
\eea 
where
\be
A_i=\cos^2{\mu}/{2}\ \Sigma_i \, ,
\ee
and,
\be
A= \cos \th \, d\ph+\sin \th(\cos \ph A_1 +\sin \ph A_2)+\cos \th A_3 \, . \label{GAU}
\ee
$\sigma_i$'s are left-invariant one-forms that are chosen as follows: 
\bea
&&\si_1= \sin \ph\, d\th +\sin \th \cos \ph \,  d\tau\, ,\nn \\
&&\si_2= -\cos \ph\, d\th +\sin \th \sin \ph\,  d\tau\, , \  \nn \\
&& \si_3= -d\ph +\cos \th\,  d\tau \, . \nn
\eea

In the new form of the metric (\ref{NEE}), we can further rescale the $U(1)$ fibers so that 
the Ricci tensor (in a basis we introduce shortly) is still diagonal. Hence, we take the metric to be
\bea
ds^2_{S^7}&=& d\mu^2 +\f{1}{4} \sin^2 \mu\, \Sigma_i^2 +\la^2 (d\th -\sin \ph A_1+\cos \ph A_2)^2 \nn \\
&+&\la^2 \sin^2 \th\, (d\ph -\cot \th(\cos \ph A_1 +\sin \ph A_2)+A_3)^2 +\lat^2(d\tau-A)^2\, ,\label{NE}
\eea 
and choose the following basis
\bea
&& e^0=\, d\mu\, , \ \ \ \ e^i=\f{1}{2} \sin \mu\, \Sigma_i\, ,\nn \\ 
&& e^5=\la (d\th -\sin \ph A_1+\cos \ph A_2) \, ,\nn \\
&& e^6=\la \sin \th(d\ph -\cot \th(\cos \ph A_1 +\sin \ph A_2)+A_3)\, ,\nn \\
&& e^7=\lat (d\tau-A)\, . \label{VI}
\eea
In this basis the Ricci tensor is diagonal and reads
\bea
&& R_{00}=R_{11}=R_{22}=R_{33}=3-\la^2-{\lat^2}/{2} \, ,\nn \\
&& R_{55}=R_{66}=\la^2+{1}/{\la^2}-{\lat^2}/{2\la^4}\, ,\ \ \ \ \ R_{77}=\lat^2+{\lat^2}/{2\la^4}\, .\label{EE}
\eea 

\subsection{The ansatz}
As in the previous section, a natural 3-form to begin with is $\om_3=e^{567}$. To proceed, however, it proves useful 
to define the following forms
\bea
R_1&=& \sin \ph (e^{01}+e^{23}) -\cos \ph(e^{02}+e^{31}) \, ,\nn \\
R_2&=& \cos \th\cos \ph (e^{01}+e^{23})+ \cos \th \sin \ph (e^{02}+e^{31})-\sin \th (e^{03}+e^{12})\, ,\nn \\
K&=&  \sin \th\cos \ph (e^{01}+e^{23})+ \sin \th \sin \ph (e^{02}+e^{31})+\cos \th (e^{03}+e^{12})\, .\label{THREE}
\eea
The key feature of this definition, that we will use frequently in this paper, is that these three forms are orthogonal 
to each other, i.e.,
\be
R_1\we R_2=K\we R_1=K\we R_2=0\, .
\ee
Let us also define,
\be
{\rm Re}\, \Omega =R_1\we e^5 + R_2 \we e^6 \, ,\ \ \ \ {\rm Im}\, \Omega =R_1\we e^6 - R_2 \we e^5\, ,
\ee
we will further need to work out the exterior derivatives of the above forms
\be
d\ro= 4\la\om_4-\f{2}{\la}\, e^{56}\we K\, ,\ \ \ \  d\io =0 \, ,\label{OM}
\ee
for $d\om_3$ in the new basis we get 
\be
d\om_3=\la\, \io\we e^7-\lat\, e^{56}\we F\, ,
\ee
with
\be
F=dA =- K -e^{56}/{\la^2}\, .\label{F} 
\ee

Note that since 
\be
d\io =0\, ,\ \ \ \ \ \io \we F =-\io\we K=0\, ,
\ee
we have three independent 4-forms $\om_4$, $e^7\we \io$, and $e^{56}\we K$, which are closed and do not contract 
into each other. Furthermore, the set of these 4-forms is closed under $d*$ operation, and hence a suitable ansatz for $F_4$ is as follows
\be
F_4=N\ep_4 + \al\, \om_4+\bet\, e^7\we \io +\ga\, K\we e^{56}\, ,\label{ans}
\ee
for $\al$, $\bet$, and $\ga$ three real constants. Taking the Hodge dual we have
\be
*_{11}F_4=N\om_3\we \om_4 +\ep_4\we (\al\, \om_3-\bet\, \ro +\ga\, K\we e^{7})\, .
\ee
Using
\be
de^{56}=\la\, \io\, ,\ \ \ \ dK=-\f{1}{\la}\, \io\, ,\label{EK}
\ee
and (\ref{OM}), we see that Maxwell equations (\ref{MAX}) reduce to 
\bea
&& -\al\la^2 +N\la \bet +\ga=0\, ,\nn \\
&& \al\lat +{2}\bet/\la+(\lat/\la^2+N)\ga=0\, ,\nn \\
&& N\al-4\la\bet+2\lat\ga=0\, .\label{M}
\eea
As for the Einstein equations, we use (\ref{EE}) and the ansatz (\ref{ans}) to obtain
\bea
&& 3-\la^2-\f{\lat^2}{2}=\f{1}{3}(\al^2+\bet^2+\f{1}{2}\ga^2+ \f{1}{2}N^2)\, ,\nn \\
&& \la^2+\f{1}{\la^2}-\f{\lat^2}{2\la^4}=\f{1}{3}(-\f{\al^2}{2}+\bet^2+{2}\ga^2+ \f{1}{2}N^2)\, ,\nn \\
&& \lat^2+\f{\lat^2}{2\la^4}=\f{1}{3}(-\f{\al^2}{2}+4\bet^2-\ga^2+ \f{1}{2}N^2)\, .\label{E}
\eea 

In general, it is not easy to solve set of coupled equations (\ref{M}) and (\ref{E}). In fact, apart from 
the known solutions, we have found no real (i.e., real coefficients for $F_4$) solutions. In especial cases, though,  
we can reduce the equations further and find solutions. Let us start by assuming
\[
\la=\lat\, ,
\]
then by the Einstein equations we must have $\bet^2=\ga^2$. Taking $\bet=-\ga$ yields $\la=\lat=1/\sqrt{5}$, $N=-6/\sqrt{5}$, 
and $\al^2=\bet^2=\ga^2=9/5$ which is the squashed solution (with torsion) of the previous section with  $R_{\mu\nu}=-{45}/{10}\, g_{\mu\nu}$. 

For $\bet=\ga$, we get $\la =\lat =1$, $N=-2$, and
$\al^2=\bet^2=\ga^2=1$; this is an Englert type solution with $R_{\mu\nu}=-{5}/{2}\, g_{\mu\nu}$. This has 
the same four-dimensional Ricci tensor as the original solution found by Englert in \cite{ENG} using parallelizing  torsions on the 7-sphere, and later by \cite{DUF3} and \cite{WAR} using Killing spinors.

\subsection{Pope-Warner solution}
In this section we rederive the Pope-Warner ansatz and the solution \cite{WAR} using the canonical forms language. 
Let us then begin by defining
\be
{\rm Re}\, L = -R_1\we e^5 + R_2 \we e^6 \, ,\ \ \ \ \ \ \
{\rm Im}\, L =R_1\we e^6 + R_2 \we e^5\, .\label{L}
\ee
We note that in the vielbein basis (\ref{VI}), $A$ in (\ref{GAU}) can be written as 
\be
A=\cot \th\, \f{e^6}{\la}+\f{\cot \mu /2}{\sin \th}\, (\cos \ph\, e^1 +\sin \ph\, e^2)\, ,
\ee
which, together with (\ref{THREE}), allows us to write $de^5$ and $de^6$ more compactly as  
\be
de^5=-e^6 \we A +\la R_1\, , \ \ \ \ \ de^6=e^5\we A +\la R_2\, .\label{56}
\ee
Taking the exterior derivative once more yields
\be
\la dR_1=\la R_2\we A +e^6\we K\, , \ \ \ \ \  \la dR_2=-\la R_1 \we A -e^5\we K\, .\label{RR}
\ee
Having derived (\ref{56}) and (\ref{RR}), it is now easy to prove that
\be
d {\rm Re}\, L =-2 A\we {\rm Im}\, L\, , \ \ \ \ \ d {\rm Im}\, L =2 A\we {\rm Re}\, L\, .\label{LL}
\ee

To absorb $A$ into $e^7$ in the above equations, we define 
\be
P=e^{-2i\tau} L\, ,\label{P}
\ee
by using eqs. (\ref{LL}), we see that
\be
dP=-\f{2i}{\lat}\, e^7\we P  \, .\label{PP}
\ee
On the other hand, note that 
\be
*L=i L\we e^7\, ,
\ee
so we can write (\ref{PP}) as
\be
dP=\f{2}{\lat}\,* P  \, .\label{PP2}
\ee
This implies that for the 4-form field strength we can take  
\be
F_4=N\ep_4  +\et\, e^7\we\, (\sin 2\tau\, {\rm Re}\, L -\cos 2 \tau\, {\rm Im}\, L)\, ,
\ee
with $\et$ a real constant. Maxwell eq. (\ref{MAX}) then requires $N=-2/\lat $, whereas, 
the Einstein equations imply $\la^2=1$, and $\lat^2=2$, together with $\et^2=2$. 
Note that in this solution the $U(1)$ fibers of $S^7$ are stretched by a factor of ${2}$. 

We can construct another consistent ansatz by taking a linear combination of Pope-Warner ansatz and 
the one introduced in the previous section. However, by this we get non-zero off diagonal components of 
energy-momentum tensor, i.e., $T_{56}\neq 0$, unless we set $\bet =0$. Let us then set
\be
F_4=N\ep_4 + \al\, \om_4 +\ga\, K\we e^{56} +\et\, e^7\we\, (\sin 2\tau\, {\rm Re}\, L -\cos 2 \tau\, {\rm Im}\, L)\, ,
\label{WW}
\ee
Maxwell eqs. (\ref{MAX}) and (\ref{M}) then require
\be
N=-2/\lat\, ,\ \ \ \ \la^2=\lat^2=1\, ,\ \ \ \ \al =\ga\, ,
\ee 
while, the Einstein equations imply
\be
\al^2=\ga^2=\et^2=1\, , \ \ \ \ \ 
\ee
which is the Englert solution with $R_{\mu\nu}=-{5}/{2}\, g_{\mu\nu}$. Note that here we have $\al=\ga=1$, 
hence the second and the third terms in 
(\ref{WW}) combine to
\be
\om_4 +\, K\we e^{56} = \f{1}{2}\, F\we F\, ,\label{FWF}
\ee 
with $F$ the K\" ahler form defined in (\ref{F}). We can now recognize (\ref{WW}) as exactly the Englert solution of  \cite{WAR}. The $F\we F$ term and the term proportional to $\et$ are each invariant under an $SU(4)$ symmetry, but 
with the given values of the constant coefficients, $\al ,\bet $, and $\ga$, the symmetry enhances to $SO(7)$.

\section{A new $AdS_5\times{\bf CP}^3$ compactification} 
With the ansatz introduced in Sec. 3.1, we can think of eleven dimensional metrics which 
are direct product of 5 and 6-dimensional spaces with $F_4$ given by (\ref{ans}) setting $N$ and $\bet$ 
equal to zero. By this, apart from the result of \cite{PN} we derive a new solution of $AdS_5\times {\bf CP}^3$ so that 
the ${\bf CP}^3$ factor is stretched along its $S^2$ fiber by a factor of $2$. 

Let us then take the eleven dimensional spacetime to be the direct product of a 5 and 6-dimensional 
spaces, 
\be
ds_{11}^2=ds_{5}^2 +ds_{6}^2\, .
\ee
For the 6-dimensional space we take the same metric that appeared in $S^7$ description in (\ref{NEE}):
\bea
ds^2_{6}&=& d\mu^2 +\f{1}{4} \sin^2 \mu\, \Sigma_i^2 +\la^2 (d\th -\sin \ph A_1+\cos \ph A_2)^2 \nn \\
&+&\la^2 \sin^2 \th\, (d\ph -\cot \th(\cos \ph A_1 +\sin \ph A_2)+A_3)^2\, , \label{NEE2}
\eea 
as mentioned before, this is an $S^2$ bundle over $S^4$, and for $\la^2=1$ we get the Fubini-Study metric on 
${\bf CP}^3$.
By taking the basis $e^0,\dots ,e^6$ as in (\ref{VI}) the Ricci tensor reads
\bea
&& R_{00}=R_{11}=R_{22}=R_{33}=3-\la^2 \, ,\nn \\
&& R_{55}=R_{66}=\la^2+{1}/{\la^2}\, \, .\label{EE2}
\eea 

As for $F_4$, we choose the following ansatz
\be
F_4=\al\, \om_4  +\ga\, K\we e^{56}\, ,
\ee
which is closed. Taking the Hodge dual we have
\be
*_{11}F_4=\ep_5\we\, (\al\, e^{56} +\ga\, K)\, .
\ee
As $F_4\we F_4=0$, in this case the Maxwell equation reads
\be
d*_{11}F_4=-(\al\, \la- \ga\, /\la)\, \ep_5\we \io =0\, ,
\ee
where use has been made of (\ref{EK}). So, we must have
\be
\al \la^2=\ga\, .\label{LA}
\ee

The Einstein equations along compact 6 dimensions, on the other hand, imply
\bea
&& 3-\la^2=\f{1}{3}(\al^2+\f{1}{2}\ga^2)=\f{1}{3}(1+\f{\la^4}{2})\al^2 \, ,\nn \\
&& \la^2+\f{1}{\la^2}=\f{1}{3}(-\f{\al^2}{2}+{2}\ga^2)=\f{1}{3}(-\f{1}{2}+{2}{\la^4})\al^2\, ,
\eea 
where we used (\ref{LA}) in the last equalities. From the above equations we get two solutions:
\be
\la^2=1\, ,\ \ \ \ \ \al^2=\ga^2=4\, ,\label{FFF}
\ee
for which the metric is the standard Fubini-Study metric of ${\bf CP}^3$. The 5d Ricci tensor becomes
\be
R_{\mu\nu}=-2\, g_{\mu\nu}\, ,   
\ee
with $\mu ,\nu =0,\ldots, 4$. Therefore the 5-dimensional spacetime is anti-de Sitter. This solution was first 
derived in \cite{PN}.

For the second solution we have
\be
\la^2=2\, ,\ \ \ \ \al^2=1\, ,  \ \ \ \ga^2=4\, ,\label{SS}
\ee
with the 5d Ricci tensor; 
\be
R_{\mu\nu}=-\f{3}{2}\, g_{\mu\nu}\, .
\ee
This new solution corresponds to an stretched ${\bf CP}^3$, in which the $S^2$ fibers are stretched by 
a factor of $2$. Note that, for this solution the 6-dimensional metric is no longer Einstein. 
Also, note that according to our discussion at the end of the previous section the first solution, (\ref{FFF}), has an 
$SU(4)$ symmetry, whereas in the new solution, (\ref{SS}), this symmetry is reduced to $SO(3)\times SO(5)$, i.e, to 
the direct product of the symmetry subgroups of the fiber and the base.

\section{$AdS_2\times H^2 \times S^7$ compactification} 
With metric (\ref{NE}) for the $S^7$, we can take yet another ansatz for the metric 
and $F_4$ and come up with a new compactification. In fact, in this section we obtain a new solution of 
type $AdS_2\times H^2 \times S^7$, with $H^2$ a hyperbolic surface. As we will see, this solution exists only 
in 11-dimensional space with Euclidean signature, and like the Pope-Warner solution the $S^7$ metric gets 
stretched along its $U(1)$ fibers by a factor of $2$. 

Let the eleven dimensional spacetime to be the direct product of two 2-dimensional spaces and $S^7$, 
\be
ds_{11}^2=ds_{A}^2 +ds_{2}^2 +ds_{S^7}^2\, ,
\ee
where $ds_{S^7}^2$ is the same as (\ref{NE}). For $F_4$ we take
\be
F_4=N\ep_2^A\we \ep_2 + \al\, \om_4+\bet\, e^7\we \io +\ga\, K\we e^{56} +\ep_2\, \we\, (\xi_1 K+\et_1 e^{56}) 
+\ep_2^A\, \we\, (\xi_2 K+\et_2 e^{56})\, ,\label{FF}
\ee
note that the first four terms are the same as those appeared in (\ref{ans}). $\xi_1, \xi_2, \et_1$, and $\et_2$ 
are constant parameters. We take the 4-dimensional space to be the direct product of two Euclidean subspaces with 
$\ep_2^A$ and $\ep_2$ as their 2-dimensional volume elements.

The Bianchi identity requires that
\be
\xi_1=\la^2 \et_1\, , \ \ \ \ \ \xi_2=\la^2 \et_2\, ,\label{xii}
\ee
For the Maxwell equation, first note that in Euclidean 11-dimensional space we need to account for an extra $i$ factor 
coming from the Chern-Simons term so that (\ref{MAX}) is replaced by
\be
d*_{11}F_4=-\f{i}{2}\, F_4\we\, F_4 \, ,\label{EUC}
\ee
therefore, with our ans\"atze (\ref{NE}) and (\ref{FF}) the Maxwell equations reduce to the following algebraic equations
\bea
&&\lat\, (2\xi_1+\et_1/\la^2)=-i(2\xi_2\ga+\al \et_2) \, ,\nn \\
&& \lat\, (2\xi_2+\et_2/\la^2)=-i(2\xi_1\ga+\al \et_1)\, , \nn \\
&& \al\lat+2\bet/\la  +\ga(\lat/\la^2 +i N)=-i(\xi_1\et_2+\et_1\xi_2)\, ,\nn \\
&& i N\al -4\la\bet +2\lat \ga=-2i\xi_1\xi_2 \, ,\nn \\
&& -\al \la^2+i N\la\bet +\ga=0\, . \label{FOUR}
\eea
Using (\ref{xii}), the first two equations above imply 
\be
\xi_1^2=\xi_2^2\, ,\ \ \ \ \ \ \et_1^2=\et_2^2\, .\label{ETA}
\ee
Had we chosen a Lorentzian signature metric for the 4-dimensional space, since $*\ep_2^A=-\ep_2$ 
and $*\ep_2=\ep_2^A$  we would have obtained $\xi_1^2=-\xi_2^2$, with no real solution. On the other hand, in 
Euclidean 11-dimensional space eq. (\ref{EUC}) implies that whenever the RHS is nonvanishing $F_4$ is necessarily 
complex valued, and so there is no restriction on the coefficients of $F_4$ to be real. However, for having a  well-defined 
metric we still require that $\la$ and $\lat$ to be real.
 
To carry on, we set $\xi_1=\xi_2\equiv \xi$ without loss of generality, and (\ref{FOUR}) becomes
\bea
&&\al \la^2 +2\la^4\ga -i\lat (1+2\la^4)=0 \, ,\nn \\
&& (\lat\la^2-i N)\, \al+6\la\bet+ (i N\la^2-\lat)\ga=0 \, ,\nn \\
&& \lat \ga-2\la \bet+i N\al/2+i\xi^2=0\, , \nn \\
&& -\al \la^2+i N\la\bet +\ga=0\, , \label{M1} 
\eea
where the second equation is obtained by dividing the third and fourth equations in (\ref{FOUR}). 
For $\bet\neq 0$, we have found no solution of (\ref{M1}) for which $\la$ and $\lat$ are both real. Let us then 
discuss the case with $\bet =0$. In this case, the second and the forth equations above imply
\be
\la^2=1\, ,\ \ \ \ \ \al=\ga\, .
\ee
Plugging this into the first equation we have
\be
\al=\ga=i \lat\, ,\label{W1}
\ee
and finally the third equation gives
\be
\xi^2=-\lat^2-\f{i}{2}\lat N\, .\label{xx}
\ee 

Let us now look at the Einstein equations along $S^7$. Taking into account 
$\la^2=1$ and $\al=\ga$, they read
\bea
&& 2-\f{\lat^2}{2}=\f{1}{2}\al^2- \f{1}{6}N^2 -\f{1}{2}\xi^2\, ,\nn \\
&& \f{3}{2}\, \lat^2=-\f{1}{2}\al^2- \f{1}{6}N^2 -\f{1}{2}\xi^2\, ,
\eea 
note the sign change of $N^2$ as a result of using the Riemannian signature (compare with (\ref{E})). Using (\ref{W1}), we can solve for $\lat$:
\be
\lat^2=2\, ,\label{W2}
\ee
and,
\be
N^2+3\xi^2+12=0\, ,\label{W3}
\ee
this last equation together with (\ref{xx}) can be solved to give $N$ and $\xi$.

The Ricci tensor along two 2-dimensional spaces reads
\bea
&&R_{ab}=(\xi^2_2+\f{1}{2}\et^2_2+\f{1}{2}N^2 -\f{1}{6}(N^2+ \al^2 +2\ga^2)-\f{1}{2}\xi^2)\, g_{ab}\, ,\nn \\
&&R_{a' b'}=(\xi^2_1+\f{1}{2}\et^2_1+\f{1}{2}N^2 -\f{1}{6}(N^2+\al^2 +2\ga^2)-\f{1}{2}\xi^2)\, g_{a' b'}\, ,
\eea
with $a,b=0,1$, and $a', b'=2,3$. Now, using (\ref{W1}), (\ref{W2}), and (\ref{W3}) , we get
\bea
&&R_{ab}=-3\, g_{ab}\, ,\nn \\
&&R_{a' b'}=-3\, g_{a' b'}\, .
\eea
Therefore, the 4-dimensional space is a direct product of a Euclidean $AdS_2$ and a 2-dimensional  hyperbolic 
surface.  Interestingly, this solution has some common features with the Pope-Warner and 
the Freund-Rubin solutions. As in the Pope-Warner solution, here the metric of $S^7$ is stretched by a factor of 
$2$ along its $U(1)$ fiber with $SU(4)$ isometry group. And, on the other hand, the 4-dimensional Ricci tensor 
is equal to that of the Freund-Rubin solution.

\section{Conclusions}
In this paper we provided a unified approach to study the squashed, stretched, and the Englert type solutions 
of 11-dimensional supergravity, especially when there are fluxes in the compact direction. With the special 
form of the metric (\ref{NE}),  we were able to construct more general ans\"atze by bringing together the 
earlier known ones and those constructed  in Sec. 3. We then used the ansatz to reduce the field equations 
to algebraic ones and rederive the known solutions. Further, using these ans\"atze we were able to find new 
compactifying solutions to 5 and 4 dimensions.  
In compactifying to 5 dimensions, we derived a solution of $AdS_5\times {\bf CP}^3$ type with the 
${\bf CP}^3$ factor stretched.  We also derived a solution of $AdS_2\times H^2\times S^7$ type compactifying 
to Euclidean 4 dimensions. In this solution, the compact space was a stretched $S^7$.   

Having derived the above solutions, the next important issue to address is that of stability. 
It is also worth studying the new solutions in the context of holographic superconductivity in 
M-theory \cite{GAU2}.


\hspace{30mm}


\vspace{1.5mm}

\noindent

\vspace{1.5mm}

\noindent

\newpage




\vspace{3mm}



\begin{thebibliography}{999}

\bibitem{RUB} P.~G.~O.~Freund and M.~A.~Rubin, {\em Dynamics Of Dimensional Reduction}, 
Phys.\ Lett.\  B {\bf 97}, 233 (1980).

\bibitem{ENG} F.~Englert, {\em Spontaneous Compactification Of Eleven-Dimensional Supergravity}, 
Phys.\ Lett.\  B {\bf 119}, 339 (1982).

\bibitem{AWA} M.~A.~Awada, M.~J.~Duff and C.~N.~Pope, {\em N = 8 supergravity breaks down to N = 1},  
Phys.\ Rev.\ Lett.\  {\bf 50}, 294 (1983).

\bibitem{DUF} M.~J.~Duff, B.~E.~W.~Nilsson and C.~N.~Pope, 
{\em Spontaneous Supersymmetry Breaking By The Squashed Seven Sphere}, 
Phys.\ Rev.\ Lett.\  {\bf 50}, 2043 (1983).

\bibitem{BAI} F.~A.~Bais, H.~Nicolai and P.~van Nieuwenhuizen, {\em Geometry Of Coset Spaces And Massless 
Modes Of The Squashed Seven Sphere In Supergravity}, Nucl.\ Phys.\  B {\bf 228}, 333 (1983).

\bibitem{DUF3} M.~J.~Duff, {\em Supergravity, The Seven Sphere, And Spontaneous Symmetry Breaking},  
Nucl.\ Phys.\ B {\bf 219}, 389 (1983).  


\bibitem{ROM} L.~Castellani and L.~J.~Romans, {\em N=3 And N=1 Supersymmetry In A New Class 
Of Solutions For D = 11 Supergravity}, Nucl.\ Phys.\  B {\bf 238}, 683 (1984).

\bibitem{SOR} D.~V.~Volkov, D.~P.~Sorokin and V.~I.~Tkach, {\em Supersymmetry Vacuum Configurations in   
D = 11 Supergravity},  JETP Lett.\  {\bf 40}, 1162 (1984).  

\bibitem{SOR2} D.~P.~Sorokin, V.~I.~Tkach and D.~V.~Volkov, {\em On the Relationship between Compactified Vacua 
of D = 11 and D = 10 Supergravities},  Phys.\ Lett.\ B {\bf 161}, 301 (1985).  

\bibitem{ENGTO} F.~Englert, M.~Rooman and P.~Spindel, {\em Supersymmetry Breaking By Torsion 
And The Ricci Flat Squashed Seven Spheres}, Phys.\ Lett.\  B {\bf 127}, 47 (1983).

\bibitem{WAR} C.~N.~Pope and N.~P.~Warner, {\em An SU(4) Invariant Compactification Of D = 11 
Supergravity On A Stretched Seven Sphere}, Phys.\ Lett.\  B {\bf 150}, 352 (1985).

\bibitem{VAS} V.~D.~Lyakhovsky and D.~V.~Vassilevich, {\em Algebraic Approach to Kaluza-Klein Models},   
Lett.\ Math.\ Phys.\  {\bf 17}, 109 (1989).  

\bibitem{GAU1} J.~P.~Gauntlett, S.~Kim, O.~Varela and D.~Waldram, {\em Consistent 
supersymmetric Kaluza-Klein truncations with massive modes}, 
JHEP {\bf 0904}, 102 (2009)  [arXiv:0901.0676 [hep-th]].  

\bibitem{CAS} D.~Cassani and P.~Koerber, {\em Tri-Sasakian consistent reduction},  
JHEP {\bf 1201}, 086 (2012)  [arXiv:1110.5327 [hep-th]].  

\bibitem{DUF2} M.~J.~Duff, B.~E.~W.~Nilsson and C.~N.~Pope, {\em Kaluza-Klein Supergravity}, 
Phys.\ Rept.\  {\bf 130}, 1 (1986).

\bibitem{HOPF} B.~E.~W.~Nilsson and C.~N.~Pope,
{\em Hopf Fibration Of Eleven-Dimensional Supergravity}, 
Class.\ Quant.\ Grav.\  {\bf 1}, 499 (1984).

\bibitem{FONT} G.~Aldazabal and A.~Font, {\em A Second look at N=1 supersymmetric AdS(4) vacua of type IIA supergravity},  JHEP {\bf 0802}, 086 (2008), [arXiv:0712.1021].  

\bibitem{PN} C.~N.~Pope and P.~van Nieuwenhuizen, {\em Compactifications of $d=11$ Supergravity on  
K\"ahler Manifolds},  Commun.\ Math.\ Phys.\  {\bf 122}, 281 (1989).  

\bibitem{GAU2} J.~P.~Gauntlett, J.~Sonner and T.~Wiseman,
{\em Quantum Criticality and Holographic Superconductors in M-theory}, JHEP {\bf 1002}, 060 (2010), 
[arXiv:0912.0512 [hep-th]].  


\end{thebibliography}
\end{document}